\documentclass[compsoc,conference,a4paper,10pt,times]{IEEEtran}
\IEEEoverridecommandlockouts

\usepackage[T1]{fontenc}
\usepackage[utf8]{inputenc}
\usepackage{adjustbox}
\usepackage{amsmath}
\usepackage{amssymb}
\usepackage{amstext}
\usepackage{amsfonts}
\usepackage{array}
\usepackage[american]{babel}
\usepackage{boldline}
\usepackage{booktabs}
\usepackage{calc}
\usepackage{cite}
\usepackage{colortbl}
\usepackage{csquotes}
\usepackage[shortlabels]{enumitem}
\usepackage{etoolbox}
\usepackage{fancyvrb}
\usepackage{filecontents}
\usepackage{float}
\usepackage{framed}
\usepackage{graphicx}
\usepackage[hidelinks]{hyperref}
\usepackage{lmodern}
\usepackage{multirow}
\usepackage[all]{nowidow}
\usepackage{pgfplotstable}
\usepackage{rotating}
\usepackage{tabularx}
\usepackage{textcomp}
\usepackage{wasysym}
\usepackage{xcolor}
\usepackage{xspace}

\usepackage{url}
\usepackage{breakurl}

\def\BibTeX{{\rm B\kern-.05em{\sc i\kern-.025em b}\kern-.08em
    T\kern-.1667em\lower.7ex\hbox{E}\kern-.125emX}}

\newcommand{\ie}{i.e.,\xspace}
\newcommand{\eg}{e.g.,\xspace}
\newcommand{\definition}[1]{\noindent\textbf{#1:}}

\begin{document}

\title{SoK: Delegation and Revocation, \\ the Missing Links in the Web's Chain of Trust}

\author{
	\IEEEauthorblockN{
		Laurent Chuat\IEEEauthorrefmark{1},
		AbdelRahman Abdou\IEEEauthorrefmark{2},
		Ralf Sasse\IEEEauthorrefmark{1},
		Christoph Sprenger\IEEEauthorrefmark{1},
		David Basin\IEEEauthorrefmark{1},
		Adrian Perrig\IEEEauthorrefmark{1}}
	\medskip
	\IEEEauthorblockA{\IEEEauthorrefmark{1}Department of Computer Science, ETH Zurich}
	\IEEEauthorblockA{\IEEEauthorrefmark{2}School of Computer Science, Carleton University}
}

\maketitle

\begin{abstract}
The ability to quickly revoke a compromised key is critical to the security of any public-key
infrastructure. Regrettably, most traditional certificate revocation schemes suffer from latency,
availability, or privacy problems. These problems are exacerbated by the lack of a native delegation
mechanism in TLS, which increasingly leads domain owners to engage in dangerous practices such as
sharing their private keys with third parties.

We analyze solutions that address the long-standing delegation and revocation shortcomings of the
web PKI, with a focus on approaches that directly affect the chain of trust (i.e., the X.509
certification path). For this purpose, we propose a 19-criteria framework for characterizing
revocation and delegation schemes. We also show that combining short-lived delegated credentials or
proxy certificates with an appropriate revocation system would solve several pressing problems.
\end{abstract}


\section{Introduction}
\label{sec:introduction}

\looseness=-1
Certificate revocation has always been a challenge in the HTTPS public-key infrastructure (or web
PKI, for short). Certificate revocation lists (CRLs) \cite{rfc5280,rfc6818} grow linearly in the
number of revocations, making their communication to browsers inefficient. CRLs and
OCSP~\cite{rfc2560} require an extra round-trip communication initiated by the browser to verify a
certificate's validity. This increases  page-load delay and reveals  users' browsing habits. The
extra round trip may also block the connection when the browser fails to receive a
response~\cite{stark2012case}. Failing open (\ie proceeding with the connection anyway) is
effectively equivalent to not checking if a certificate is revoked, jeopardizing security.

Given the above problems, some browser vendors have decided to disable online revocation checks (\ie
CRL and OCSP) and instead rely upon small sets of emergency revocations pushed to clients through
software updates~\cite{langley2012,mozilla_rev}. OCSP stapling~\cite{rfc6961} addresses the
extra-round-trip problem but, like CRLs, places an additional burden and reliance on certification
authorities (CAs) as they must be frequently contacted over a certificate's lifespan. Moreover, a
stapled certificate status is typically valid for four days, implying a rogue certificate or a
compromised private key remains usable by the adversary for four days in the worst case; the
consequences can thus be severe.

Short-lived certificates~\cite{rivest1998can,topalovic2012towards} provide comparable benefits to
OCSP stapling. Again, four days is the suggested validity period. Questions remain, however, about
the feasibility of reducing this period to several minutes to limit adversarial capabilities in case
of private key compromise. In general, the tradeoff between promptly disseminating revocation
information to browsers (requiring an extra round-trip and burdening CAs) and increasing the
system's efficiency (but sacrificing security) is now well established. Unfortunately, browser
vendors often favor efficiency over security when it comes to certificate
revocation~\cite{liu2015end}.

Besides revocation, new requirements with associated problems have emerged for the web PKI. In
particular, content delivery networks (CDNs) are now widely used and beg for a secure delegation
system~\cite{Liang:2014im}. In practice, rather than explicitly delegating specific rights to CDNs,
domain owners often resort to some form of key sharing~\cite{Cangialosi:2016jx}. The delegation
problem is, in fact, intimately related to that of revocation. If we could rely on an efficient and
secure revocation system, the negative consequences of key sharing would be minimized as compromised
keys could be invalidated as soon as misbehavior is observed.

This evolution of the HTTPS landscape has put domain owners into an uncomfortable position: to
satisfy their delegation and revocation needs, they must either adopt insecure approaches or heavily
depend on CA support. We therefore ask the following fundamental question: \emph{Why are domain
owners required to visit a CA for every issuance, renewal, revocation, and key update they perform
to any of their subdomains?} 

\looseness=-1
In our view, domain owners should be offered a flexible and secure solution that leaves them in full
control of their domain and its subdomains. We thus formulate the following two requirements for any
possible solution. Domain owners should be able to autonomously
\begin{enumerate}[label=\textbf{(R\arabic*)}, ref=R\arabic*, leftmargin=\widthof{(R1)}+\labelsep]
\item\label{req:revoke} decide on the validity period or revocation status of their own
certificates; and
\item\label{req:delegate} delegate all or a subset of their privileges to third parties, without
sharing a private key with them.
\end{enumerate}

Our systematization of knowledge has resulted from our endeavor to identify solutions satisfying the
above two requirements. We start by providing background on delegation to CDNs
(Section~\ref{sec:cdn}) to show how the web and its trust model have evolved in recent years. We
then review a wide spectrum of approaches to revocation and delegation with different features and
trade-offs (Section~\ref{sec:background}). We divide revocation schemes into four categories,
depending on which actor  provides the revocation information, and we distinguish delegation schemes
with and without key sharing. We also highlight several problems of these approaches, showing why
they fail to fully satisfy the requirements. Next, we examine different approaches that extend the
traditional chain of trust and promise to bring more security and flexibility to the HTTPS ecosystem
(Section~\ref{sec:models}). These include name constraints, short-lived certificates, proxy
certificates, and delegated credentials. We focus on the latter two, proxy certificates and
delegated credentials, as the most promising candidates for satisfying the requirements and discuss
them in more detail in Sections~\ref{sec:solution} and~\ref{sec:del_cred}.

The concept of proxy certificates was introduced more than a decade ago in the context of grid
computing~\cite{rfc3820}. Proxy certificates allow entities holding non-CA certificates to delegate
some or all of their privileges to other entities. We thoroughly investigate the challenges and
advantages of such delegation in the context of the current web PKI. Allowing such delegation would
provide several security and efficiency advantages. These include a domain owner's ability to update
and use multiple distinct keys (\eg for multiple subdomains, or for different transactions per
subdomain) much more easily and rapidly than with current practices. Domain owners can also issue
short-lived proxy certificates for added security. Perhaps more importantly, the owner's main
private key, corresponding to the public key in the CA-issued certificate, need not be involved in
online cryptographic operations, and could thus be safely stored on a disconnected (air-gapped)
device. Additionally, proxy certificates need not be logged by Certificate Transparency servers,
solving the problem of log servers disclosing private
subdomains~\cite{eskandarian2017certificate,stark2019ct}.

More recently, another promising approach called delegated credentials was developed at the IETF and
described in an Internet Draft~\cite{draft_subcerts} as a first step towards standardization. On 1
November 2019, Mozilla~\cite{mozilla_DC}, Cloudflare~\cite{cloudflare_DC}, and
Facebook~\cite{facebook_DC} announced plans to support delegated credentials. We describe and study
this new proposal in comparison to similar approaches, proxy certificates in particular.

In Section~\ref{sec:tls_consequences}, we analyze the consequences of using short-lived certificates
and investigate how TLS should handle these certificates.

In Section~\ref{sec:analysis}, we thoroughly analyze the different approaches to revocation and
delegation. We propose a 19-criteria framework, which includes criteria pertaining to delegation,
revocation, security, efficiency, and deployability. We use this framework to evaluate and compare
19 delegation schemes, revocation schemes, and related certificate features. We show how delegated
credentials and proxy certificates fit into the bigger picture and how they can be combined with
other schemes to satisfy the requirements and offer additional properties. Finally, in
Section~\ref{sec:related}, we discuss related topics, and  we draw conclusions in
Section~\ref{sec:conclusion}.

The main contributions of our work are to systematize existing work and suggest new solutions and
directions in this important problem domain:
\begin{itemize}
	\item After segmenting revocation schemes into 4 categories, we provide a detailed analysis of
	19 delegation and revocation schemes with respect to 19 evaluation criteria.
	\item As those schemes offer a wide set of complementary features, we examine how they can
	be combined so that both the requirements are met.
	\item We show how proxy certificates can be adapted to today's web, and present three use
	cases for such certificates.
	\item Finally, we point out that session resumption may undermine the security advantages
	provided by short-lived credentials, and we discuss potential workarounds.
\end{itemize}

We refer readers who are not familiar with the HTTPS ecosystem to
Durumeric et al.~\cite{durumeric2013analysis} or
Clark and van Oorschot~\cite{DBLP:conf/sp/ClarkO13} for background information.

\section{Delegation to CDNs}
\label{sec:cdn}

Content delivery networks (CDNs) pose new challenges for the web PKI, as they fundamentally alter
the trust model upon which HTTPS relies~\cite{Liang:2014im}. Every CDN is composed of a myriad of
servers that distribute content to users based on their location. The objective is usually to
increase performance and availability, but CDN vendors may also provide security-related services.
The caching servers controlled by the CDN, commonly referred to as edge servers, act as
intermediaries between clients and the origin server (controlled by the domain owner). At the
moment, this communication model is incompatible with TLS, which was specifically designed to
prevent man-in-the-middle (MitM) attacks, and may force domain owners to share their private key
with the CDN operator who then acts as a MitM.

Multiple options exist for redirecting HTTP requests to CDN servers. The most common
techniques are the following:
\begin{itemize}
	\item \textbf{Authoritative:} The CDN's name servers are defined as authoritative for the domain
	in question. This lets the CDN take full control over the resolution of the entire domain. When
	trying to contact \texttt{example.com}, the browser should obtain the IP address of one of the
	CDN's edge servers. As the redirection happens through DNS, the browser will attempt to
	establish a connection based on a valid certificate for \texttt{example.com}; therefore, the
	edge server must know the corresponding private key.
	\item \textbf{CNAME:} The redirection is done through a DNS Canonical Name (CNAME) record, which
	allows for a more fine-grained mapping as it supports the redirection of specific subdomains.
	For example, such a record could specify the following mapping:
	\begin{center}
		\texttt{s1.example.com} $\rightarrow$ \texttt{s1.example.com.cdn.net}
	\end{center}
	As in the previous case, if the browser tries to contact \texttt{s1.example.com}, then it will
	expect to see a valid certificate for it.
	\item \textbf{URL rewriting:} The URLs of specific resources (\eg images, videos, documents) are
	modified (either automatically by the web server, or manually by the domain owner) so that they
	point to CDN servers. This is the most fine-grained approach but it has drawbacks: URL rewriting
	does not support common security features that the CDN may offer, such as DDoS protection and
	web application firewalls. Delegation plays a less significant role in this scenario, as the
	browser can accept the CDN's own certificate.
\end{itemize}

We classify schemes with respect to delegation by whether it is
supported in any (possibly insecure) way, and whether it is possible
without requiring key sharing, as necessary by Requirement~\ref{req:delegate}.

\section{Dealing with Key Compromise}
\label{sec:background}

We now review a vast range of techniques developed to prevent, detect, and remedy the
compromise of private keys. The schemes we present will be compared in our systematic
analysis in Section~\ref{sec:analysis}.

\subsection{Revocation}
\label{sec:revocation}

Certificate revocation is a notoriously challenging aspect of the web PKI~\cite{liu2015end}. In
recent years, researchers have examined the question of \textit{how} revocations should be delivered
to clients, with objectives such as efficiency, deployability, and privacy in mind. For this reason,
we classify revocation schemes based on the delivery process. Specifically, we distinguish four
types of revocation schemes: \emph{client-driven}, where the browser establishes an out-of-band
connection with a revocation provider; \emph{server-driven}, where the TLS server attaches a recent
revocation status to the connection; \emph{vendor-driven}, where browser vendors push revocations to
clients through software updates; and \emph{middlebox-driven}, where network devices deliver
revocations. Our selection of schemes is not intended to be comprehensive; we instead seek to
present a few prominent examples in each category to help the reader understand and further classify
alternative schemes.

\subsubsection*{Category I: client-driven}
A traditional certificate revocation list (CRL)~\cite{rfc5280} simply contains a set of revocations
typically recorded by a CA. The distribution point of such a list can be specified within each
certificate. A client may fetch the entire list (which can grow quite large) from that distribution
point when establishing a TLS connection. OCSP~\cite{rfc2560} improves upon this design by letting
clients contact a special responder to determine the status of any specific certificate. Both
approaches suffer from latency and privacy issues. This is due to the fact that an extra connection
must be established to the CRL issuer or OCSP responder, revealing to that third party which domain
the user is visiting.

\subsubsection*{Category II: server-driven}
OCSP stapling~\cite{rfc6961} allows the web server to add a timestamped OCSP response (signed by the
corresponding CA) to the TLS handshake. Unfortunately, OCSP stapling is ineffective unless the
browser knows when to expect a stapled response, as an attacker could just not include any OCSP
status when using a revoked certificate. The must-staple extension~\cite{draft_muststaple} addresses
this issue but has yet to gain widespread support. PKISN~\cite{PKISN2016} tackles another problem:
collateral damage resulting from the revocation of the certificate of a large CA, \ie the sudden
invalidation of all certificates previously signed by this CA, which should be unaffected. This
requires all certificates to be timestamped by a verifiable log server, whereby CA certificate
revocations can be performed effective from a specific point in time, not affecting previously
issued certificates. Although the authors~\cite{PKISN2016} note that PKISN could also be deployed
with a vendor-driven model, we will consider its main deployment model as one where servers deliver
the revocation status to browsers with an OCSP-stapling-like mechanism.

\subsubsection*{Category III: vendor-driven}
We identified three main examples of vendor-driven revocation schemes. CRLSets~\cite{langley2012}
and OneCRL~\cite{mozilla_rev} are, respectively, Google's and Mozilla's effort to push a minimal set
of critical revocations to their browsers. CRLite~\cite{larisch2017crlite}, based on the same
approach of disseminating revocations through software updates, uses Bloom filters to efficiently
represent a larger number of revocations.

\subsubsection*{Category IV: middlebox-driven}
This category of schemes follows the observation that the communication and storage burden incurred
by revocations can be carried by a single device for multiple hosts. RevCast
\cite{schulman2014revcast} propagates revocations over FM radio using the RDS protocol. An
RDS-to-LAN bridge then delivers revocations to end hosts. In a similar vein, RITM~\cite{RITM2016}
relies on a network device on the client--server path to deliver revocations to end hosts by
analyzing and appending relevant information to TLS handshakes.

\subsection{(Revocable) Delegation}
\label{sec:delegation}

As we describe in Section~\ref{sec:cdn}, in most deployment models, hosting providers (such as CDNs)
need browser-accepted certificates for the domains they serve. This often leads to \emph{key
sharing}~\cite{Liang:2014im,Cangialosi:2016jx}, which can take various forms. In its simplest form,
key sharing consists of having the domain owner directly upload its private key(s) to the hosting
provider (through a web interface, for example). Alternatively, the hosting provider may use a
certificate with a \emph{subject alternative name} (SAN) list containing domain names from numerous
distinct customers. Such certificates raise a number of questions. Cangialosi et
al.~\cite{Cangialosi:2016jx} refer to these as ``cruise-liner'' certificates, and ask:

\begin{displayquote}
``Who on a cruise-liner certificate deserves access to the certificate's corresponding private key,
given that whoever has it can impersonate all others on the certificate? Who among them has the
right to revoke the certificate, if so doing potentially renders invalid a certificate the others
rely on? Cruise-liner certificates are not covered explicitly by X.509, but we can infer that, in
all likelihood, only the hosting provider has the private keys and right to revoke.''
\end{displayquote}

Because X.509 certificates do not natively support explicit delegation, researchers and
practitioners have proposed different approaches that would allow domain owners to use proxies
(CDNs, in particular) without any key sharing. Keyless SSL~\cite{DBLP:conf/trustcom/StebilaS15}
(developed by CloudFlare) splits the TLS handshake so that most of the connection establishment is
handled by edge servers, while operations requiring the domain's private key are delegated to a key
server maintained by the domain owner. Keyless SSL is compatible with both RSA and Diffie-Hellman
handshakes. In RSA mode, the key server decrypts the premaster secret (generated and encrypted by
the browser using the domain's public key) and sends it back to the edge server over an encrypted
channel. In the Diffie-Hellman case, the edge server sends a hash of parameters and nonces to the
key server, which the key server signs and returns. The protocol was analyzed in a cryptographic
model~\cite{Bhargavan:2017dx} where new attacks show that Keyless SSL, as specified for TLS~1.2,
does not meet its intended security goals. Additionally, a new design for Keyless SSL for both
TLS~1.2 and TLS~1.3 is given, together with a proof of security. In their design, session resumption
is forbidden except in special cases.

SSL splitting~\cite{LesniewskiLaas:2005ib} is an older but similar technique with the additional
guarantee that data served by the proxy server is endorsed by the origin server. This is achieved by
requiring the origin server to compute message authentication codes: for each record, the origin
server sends the MAC and a short unique identifier that the proxy server uses to look up the
corresponding payload in its local cache. Unfortunately, this approach limits the benefits of using
a CDN as it increases latency (even more so than Keyless SSL, which only affects the initial
handshake).

Liang et al.~\cite{Liang:2014im} proposed a solution, based on DANE~\cite{rfc6698}, that makes the
delegation to a CDN explicit through the name resolution process. The domain owner must add a
special TLSA record containing both its own certificate and the certificate of the CDN to its DNSSEC
records. This approach requires that a modified version of DANE be deployed, and that a browser
extension be installed. It also increases page-load delay as it requires an extra round trip during
the TLS handshake.

\subsection{Related Certificate Features}
\label{sec:cert_features}

Revocation and delegation, as we will show, are issues that can be addressed simultaneously. Instead
of relying on ad-hoc schemes to satisfy the two requirements stated in
Section~\ref{sec:introduction}, these requirements can be met using existing X.509 features, but
currently incurring major drawbacks.

The name constraints extension~\cite{rfc5280} allows CAs to issue CA certificates with a limited
scope. The constraint is specified as a fully qualified domain name and may  specify a host. For
example, as indicated in RFC~5280, both ``host.example.com'' and ``my.host.example.com'' would
satisfy the ``.example.com'' constraint, but ``example.com'' would not. Unfortunately, this
mechanism suffers from limited support and implementation issues. Liang et al.~\cite{Liang:2014im}
report that major browsers do not check name constraints, except for Firefox. Even when they check
name constraints, a limitation of the standard~\cite{rfc5280} can allow a dishonest intermediate CA
(normally restricted by a name constraint) to issue certificates for arbitrary domain names that is
not caught by the browser's enforcement.

Short-lived certificates~\cite{topalovic2012towards} reduce the attack window after a key compromise
but require CA support. This places an extra burden on those CAs; the burden increases with shorter
expiry windows, imposing minimum length of certificates'  validity periods. Self-signed certificates,
in contrast, allow domain owners to both delegate and select their attack window, but require trust
on first use (TOFU) or an authenticated, out-of-band transmission of the self-signed root certificate.

\subsection{Certificate Transparency}
\label{sec:ct}

The Certificate Transparency (CT) framework~\cite{rfc6962} was developed by Google in response to
several cases of CA compromise that resulted in the issuance of illegitimate certificates for
high-profile domains, including *.google.com~\cite{prins2011diginotar}. The objective of CT is to
make certificates publicly visible. To do so, CT relies on append-only log servers that anyone can
consult. To make the logging process verifiable, certificates are incorporated into a Merkle hash
tree, which allows log servers to produce efficient proofs of presence and consistency. When a
certificate is submitted to a log server---provided that the certificate is rooted in an accepted
trust anchor---the log will reply with a Signed Certificate Timestamp (SCT). This constitutes a
promise that the certificate is already or will be incorporated into the hash tree within a
predefined period. Chrome requires that SCTs be provided with all certificates issued after April
30, 2018~\cite{ct_requirements}.

Certificate Transparency is relevant to our discussion for several reasons: CT allows detecting
illegitimate CA actions and is thus a precious source of information for taking revocation
decisions. As an integral part of the current HTTPS ecosystem, CT must be accounted for when
proposing a new scheme, for both compatibility and efficiency reasons. Finally, as an extensive
source of certificates, CT is particularly helpful for understanding today's PKI.

\section{Overview of Trust Models}
\label{sec:models}

\begin{figure*}[h!]
	\centering
    \includegraphics[width=\textwidth]{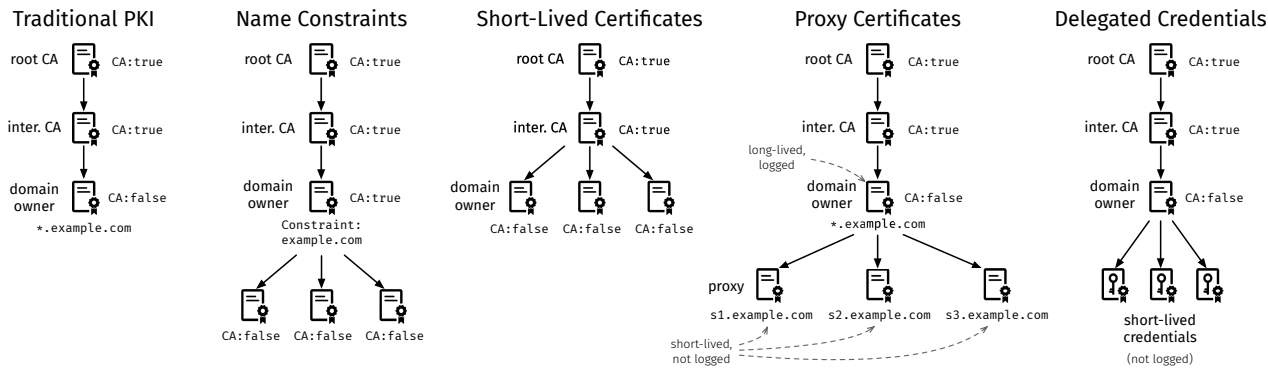}
    \caption{Illustration of different PKI models and extensions. Solid arrows represent the ``chain
    of trust''; certificates with no inbound arrow are self-signed. Only proxy certificates and
    delegated credentials may not be logged (see Section~\ref{sec:logging}).}
    \label{fig:overview}
\end{figure*}

Figure~\ref{fig:overview} illustrates the trust model of name constraints, short-lived certificates,
proxy certificates, and delegated credentials. In a traditional PKI, the ``chain of trust'' is
typically only three certificates long, and certificates are long-lived (on the order of months, if
not years). With name constraints, a link may be added to the chain in the form of a restricted CA
certificate. Theoretically, a CA certificate with the right constraints could be given to a domain
owner, but because browsers do not properly support name constraints at the moment, CAs do not offer
this option. Also illustrated in Figure~\ref{fig:overview} are classic short-lived certificates.
Previous work suggested that these should be issued by CAs~\cite{topalovic2012towards}. Short-lived
proxy certificates offer the best of both worlds: domain owners can impose their own policies
without any form of CA support. Proxy certificates allow domain owners to issue their own
certificates, as self-signed certificates do, but do not require TOFU as the certificate chain
starts from a CA certificate. With these properties, short-lived proxy certificates satisfy the
requirements (\ref{req:revoke} and \ref{req:delegate}). Delegated credentials are stripped-down
certificates, containing merely a public key and a few parameters. Similarly to proxy certificates,
delegated credentials are signed with the domain owner's private key. Moreover, the time to live of
delegated credentials must be shorter than 7 days~\cite{draft_subcerts}. 

While CA-based authentication has many advantages over trust-on-first-use or web-of-trust
approaches, proxy certificates illustrate that interactions between certificate holder and CAs may
be neither needed nor desirable for actions such as delegation or revocation. One may argue, after
the rise of Let's Encrypt, that interacting with an automated CA is free and effortless. However,
Let's Encrypt is a unique case; it is actually a costly operation supported by a large number of
sponsors. Interactions with a CA incur costs, and for Let's Encrypt to continue to thrive, the
burden put on it by new security schemes should be minimal. Moreover, each CA is a single point of
failure. While a CA being unavailable or going bankrupt may not be an issue for a domain owner
wanting to obtain a new certificate (as they can simply pick another CA), it is a problem with
regard to delegation and revocation. Finally, even if a CA provides a feature such as short-lived
certificates, it will not necessarily be flexible (the validity period may be fixed, for example).

With proxy certificates or delegated credentials, domain owners can become issuers, which raises the
question of whether they would need to fundamentally become CAs. This would imply following best
practices, using sophisticated software and hardware, and being accountable in case of a security
blunder. This is true only to the extent that the responsibilities and security requirements of
domain owners are commensurate with the scale of their endeavors: the administrator of a small
website could choose not to use any delegation mechanism at all, while an online bank may want to
spend a significant amount of time, effort, and money on protecting its private keys. In the worst
case, a failure to adequately issue credentials can only cause harm to the domain in question.

CA support is the main feature distinguishing proxy certificates from name-constraint certificates.
While certificates that specify appropriate name constraints must be obtained from CAs, proxy
certificates can be issued without interacting with a CA. Another important aspect, relating to the
above discussion, is that name constraints---because the CA bit must be set to true in the
corresponding certificate for them to be effective---reinforce the misconception that domain owners
would indispensably need to become CAs if they want to issue certificates for their own domain.
Delegated credentials also require special certificate extensions, but they do not require the CA
bit to be set to true. Delegated credentials also have the advantage of requiring minimal code
changes, compared to approaches that rely on full-fledged certificates, because of the restricted
semantics of the credentials. Unfortunately, this also implies that delegated credentials do not
give domain owners the opportunity to enforce fine-grained security policies.

\section{Proxy Certificates for the Web}
\label{sec:solution}

The concept of proxy certificates was first proposed in the early 2000s in the context of grid
computing~\cite{welch2004x, rfc3820}, where the need for secure SSL/TLS delegation emerged. Back
then, proxy certificates were used in middleware libraries and for use cases such as single sign-on.
The web was not envisioned as the main application for proxy certificates. At the time, the
technologies used on the web were vastly different: the very first CDNs had just been created,
Certificate Transparency did not yet exist, and the use of HTTPS was far less common than it is
today. In this section, we show how proxy certificates could be adapted to today's web to fulfill
the requirements.

To delegate rights to a third party, a proxy certificate is signed using the domain owner's private
key. It informs clients that the holder of the proxy certificate may legitimately serve content for
the specified domain name. The term ``proxy'' must be interpreted as an agent to which the domain
owner has conferred certain rights. Here, proxies are HTTPS servers, under the control of either the
domain owner or a third party such as a CDN or hosting provider.

As we describe in detail in Section~\ref{sec:use_cases}, several use cases exist where proxy
certificates would be beneficial. An overarching principle guiding their use is that the scope and
exposure of private keys should be limited. User-facing web servers, in particular, should only hold
private keys with limited capabilities, because they are especially exposed to various threats. This
has been demonstrated by recent vulnerabilities, such as Heartbleed~\cite{durumeric2014matter},
which allow attackers to remotely read protected memory on the vulnerable machine and thus extract
private keys.

Another important aspect to consider is the economic consequences proxy certificates would have on
the current PKI ecosystem. They could reduce the number of requests to CAs (as having a separate
certificate for each subdomain would not be necessary anymore, for example) and thus reduce their
revenue. Although this is not in the interest of CAs, they will not be able to hinder the deployment
of proxy certificates, as they are not involved in the issuance process. Reducing CA-incurred costs
is beneficial for the deployment of HTTPS, as demonstrated by the success of Let's Encrypt, for
example.

In the remainder of this paper, we use the term \emph{end-entity certificate} to refer exclusively
to a non-CA certificate issued by a CA, as opposed to \emph{proxy certificate}, which refers to a
certificate that extends the chain of trust starting from an end-entity certificate. We use the term
\emph{end-entity key} to refer to the private key that corresponds to an end-entity certificate.

\subsection{Certification Path Validation}
\label{sec:x509_validation}

A proxy certificate is similar to any other X.509 certificate in terms of format; the distinction
from a regular certificate comes from the certification path. In a nutshell, a proxy certificate is
considered valid for a given name if it extends the chain from a non-CA certificate that is valid
for that name and the proxy certificate also contains that name in its (possibly reduced) set of
permitted names. However, we also allow for chains of proxy certificates and the restriction thereof
through ``path length'' constraints, which---among other factors---makes the validation of proxy
certificates non-trivial. Below we describe a validation algorithm designed to require only minor
changes to the current X.509 specification and implementations.

An algorithm for the validation of a regular certification path is given in RFC 5280~\cite[Section
6]{rfc5280}. Trust anchor information (typically in the form of self-signed certificates) must be
provided as input to the algorithm. A prospective certification path of length $n$ (which does not
contain the trust anchor) is also provided as input, along with other information such as the
current date/time and policies. The algorithm then iterates over the $n$ certificates and verifies
that they satisfy a number of conditions. We extend that algorithm as follows:

\begin{enumerate}
	\item Split the certification path into a \emph{regular path} (stopping at the first non-CA
	certificate, \ie the end-entity certificate) and a \emph{proxy path} (consisting of zero or more
	proxy certificates).
	\item Run the algorithm exactly as specified in RFC 5280 on the regular path. If the algorithm
	indicates a failure, then stop and return the failure indication; otherwise, continue with the
	next step.
	\item If the proxy path is null, then return a success indication; otherwise, run the algorithm
	of RFC 5280 on the proxy path as follows:
	\begin{itemize}
		\item Provide the end-entity certificate as the ``trust anchor information'' input to the
		algorithm.
		\item Ignore the CA boolean in the ``basic constraints'' extension of the proxy
		certificate(s).
		\item Consider ``subject alternative name'' values in both the end-entity certificate and
		subsequent proxy certificates as additional name constraints. That is, in each iteration $i$
		of the algorithm, update the ``permitted subtrees'' state variable, which defines a set of
		names ``within which all subject names in subsequent certificates in the certification path
		MUST fall.''~\cite{rfc5280}, as follows:
		
		\smallskip
		\begin{center}
			PST$_i$ = PST$_{i-1}$ $\cap$ NC$_i$ $\cap$ (SAN$_i$ $\cup$ CN$_i$),
		\end{center}
		\smallskip
		
		where PST$_{i-1}$ is the previous value of the ``permitted subtrees'' variable, NC$_i$ is
		the set of acceptable names defined by the ``name constraints'' extension, SAN$_i$ is the
		set of names indicated in the ``subject alternative names'' extension, and CN$_i$ is the
		``common name'' field of the $i$-th certificate. Initially, PST$_0=$ SAN$_0$ $\cup$ CN$_0$,
		where SAN$_0$, and CN$_0$ are defined in the end-entity certificate.
	\end{itemize}
\end{enumerate}

The path length is interpreted as is done in RFC~5280 for the proxy path as well,
restarting the count. The above validation algorithm then guarantees that the following properties
are satisfied:

\begin{itemize}
	\item The set of acceptable names cannot be extended by an additional proxy certificate down the
	certification path. However, anyone holding a proxy certificate and the corresponding private
	key may sign another proxy certificate with an equal or smaller set of acceptable names,
	provided that the path length constraint is respected.
	\item Since every certificate in the validation path must be valid, the expiration time of a
	proxy certificate cannot be deferred by extending the validation path with an additional proxy
	certificate.
	\item The end-entity certificate must be issued by a trusted CA. It may be used as is by the
    domain owner, without any proxy certificate in the chain (\ie with a null proxy path).
\end{itemize}

A CDN holding a proxy certificate for \texttt{www.ex\-am\-ple.com} may issue a valid proxy
certificate (one level below in the chain of trust) for \texttt{www.ex\-am\-ple.com}, but not for
\texttt{ad\-min.ex\-am\-ple.com}. Note also that a wildcard certificate for \texttt{*.example.com}
would match \texttt{foo.example.com} but not \texttt{bar.foo.ex\-am\-ple.com} as per current
standards~\cite{rfc6125}. Therefore, someone holding such a wildcard certificate cannot issue valid
proxy certificates for names that would otherwise not be covered by the wildcard certificate itself,
even when extending the proxy path.

RFC 3820~\cite{rfc3820} provides an analogous logic. However, we base our algorithm upon the more
recent and complete RFC 5280~\cite{rfc5280}, which does not cover proxy certificates. In contrast to
our algorithm, RFC 3820 (a) forbids usage of the ``subject alternative name'' extension in proxy
certificates, (b) does not specify how ``name constraints'' should be treated, and (c) introduces an
additional field for restricting the length of the proxy path, rather than utilizing the existing
``path length'' parameter as we do here.

\subsection{Certificate Logging}
\label{sec:logging}

As indicated in Figure~\ref{fig:overview}, we submit that proxy certificates should not be logged
(by CT servers, for example). Our rationale is as follows: First, the CT framework was created with
the objective of uncovering CA misbehavior and compromise, not attacks against individual domain
owners. Second, comparing with the current situation, not logging proxy certificates does not reduce
security: an attacker who compromises a private key can already impersonate the corresponding
domain; issuing bogus proxy certificates for that domain would not give the attacker any more
capabilities. Therefore, logging proxy certificates does not help domain owners discover breaches.
Finally, traditional short-lived certificates are issued by CAs and, except for their validity
period, cannot be distinguished from regular certificates. This implies that each short-lived
certificate must be logged as any other certificate, increasing the pressure on log servers.
Although potential solutions have been proposed, there is no consensus on how log servers should
deal with short-lived certificates~\cite{ietf100Lopez}. If proxy certificates are not logged, they
do not suffer from this issue.

\subsection{Use Cases}
\label{sec:use_cases}

In this section, we describe three use cases that highlight the benefits proxy certificates would
bring to the web PKI. These use cases are non-exhaustive, and the features we present can be
combined in different ways. For example, a domain owner may want to (a) use a CDN, (b) have multiple
subdomains, and (c) be able to specify different policies on each subdomain.

\subsubsection*{Use Case 1: Content Delivery}

The primary use case we envision for proxy certificates is content delivery, with a potentially
large number of caching servers distributing content fetched from an origin server. The
infrastructure needed to best take advantage of proxy certificates in this scenario depends on
different factors, including whether the domain owner decides to focus on security,
deployability, or efficiency. An entire range of possible configurations exists. At one end of the
spectrum, the administrator of a static website with no sensitive data could choose to forgo using
proxy certificates altogether, as they are not mandatory. At the other extreme, the
end-entity key can be stored on an air-gapped device, kept in a secure location and configured only
to sign very-short-lived proxy certificates, which could then be extracted from the signing device
using QR codes~\cite{MatStePer2016castle} displayed on an attached screen and read by a networked
camera, for example. This might be the solution of choice for an organization with high security
requirements, such as a financial institution.

Before an HTTPS connection can be established, the client must obtain the IP address of the edge
server through DNS resolution. The edge server could be one of many servers under the control of a
CDN, for example, or a machine that the domain owner controls directly. If the edge server is
controlled by the domain owner, the DNS resolution is straightforward. Redirection to a CDN server,
on the other hand, can be realized in different ways, as described in Section~\ref{sec:cdn}.

\begin{figure}
	\includegraphics[width=\columnwidth]{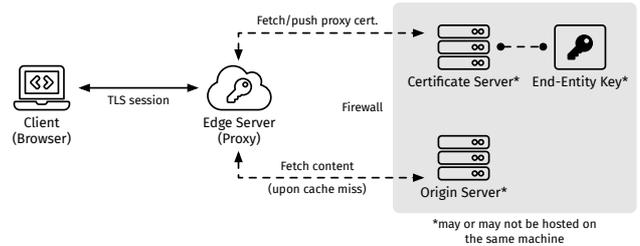}
	\caption{Deployment model for Use Case 1 (content delivery). The white key is
	certified by the end-entity key (in black) through proxy certificates.}
	\label{fig:deployment}
\end{figure}

The client then establishes an HTTPS connection with the edge server, which temporarily stores some
or all of the resources fetched from an origin server, as illustrated in
Figure~\ref{fig:deployment}. To establish a TLS connection with the client, the edge server needs a
proxy certificate that it can either fetch from the certificate server, or that the certificate
server can push at regular intervals. Issuing these proxy certificates requires an initial setup:
\begin{enumerate}
	\item The edge server (or the entity controlling it) generates a key pair and puts the public
	key into an unsigned certificate signing request (CSR) with the same format as a standard CSR.
	We call this file a \emph{proxy CSR}.
	\item The domain owner obtains the proxy CSR through an authentic channel (\eg the web portal of
	the CDN).
	\item The domain owner configures the certificate server to issue proxy certificates (based on
	the proxy CSR) at a specified frequency and with a specified validity period. Only the
	\texttt{notBefore} and \texttt{notAfter} fields~\cite{rfc5280} are updated. The validity periods
	of two consecutive proxy certificates must overlap to avoid downtimes.
\end{enumerate}

Thereafter, when needed, the certificate server creates a new proxy certificate and signs it using
the end-entity key. This implies that the certificate server must either have direct access to the
end-entity key or access to a device able to produce the appropriate signatures. For example, the
end-entity key could be stored on a hardware security module (HSM). The proxy certificate is then
transmitted from the certificate server to the edge server. As proxy certificates are public
information, any communication protocol can be employed. The CDN could even extend the certification
path further (by additional proxy certificates) so that different keys are used by
different edge servers.

Under normal circumstances, even if the proxy certificates are short-lived, the key pair used by the
edge server can remain unchanged over a long time. Indeed, the reissuance process only serves the
purpose of extending a lease to the proxy, but does not imply a key rollover. If needed, key
rollover can be achieved simply by having the edge server generate a new key pair, and use the new
public key for subsequent key signing requests. The domain owner has the option of terminating that
lease at any time, by simply stopping the issuance process, thus effectively revoking delegation
privileges. It is recommended, in such a context, to restrict traffic to the certificate and origin
servers, especially from the public Internet. A firewall may be configured to limit
incoming connections to those from the CDN's IP space, which is generally public
information~\cite{cloudflare_ips}.

Since 2016, Let's Encrypt has been offering free certificates to domain owners, thanks in part to a
completely automated domain validation process supported by the Automatic Certificate Management
Environment (ACME) protocol~\cite{rfc8555}.
Since then, several other CAs have followed suit and started using
ACME~\cite{digicert_acme,sectigo_acme}.
These CAs generally also offer tools for (re)issuing and revoking certificates.
In 2018, ACME v2 was adopted by Let's Encrypt, notably to support wildcard
certificates~\cite{acmev2}. Similar automation could
be adopted by domain owners to handle and maintain
proxy certificates, which could be implemented using a software framework like that of Let's
Encrypt, or incorporated into it.

\subsubsection*{Use Case 2: Private, Separated Subdomains}

Another use case is that of using different private keys on different subdomains. Wildcard
certificates (albeit generally more expensive than regular ones) are appreciated by domain owners
because they allow them to protect any number of subdomains they want, independently from CAs.
However, a wildcard certificate normally implies that the same private key is used on all
subdomains. In contrast, by using a wildcard in the end-entity certificate (such as
\texttt{*.example.com}) but a more specific name (such as \texttt{s1.example.com}) in all proxy
certificates (see Figure~\ref{fig:overview}), the domain owner can make sure that the consequences
of a key compromise or misbehavior from a hosting provider are confined to the corresponding
subdomain.

Moreover, as proxy certificates need not be included in certificate logs such as CT's (see
Section~\ref{sec:logging} above for motivations), they would not disclose private subdomain names
such as \texttt{se\-cret-pro\-ject.ex\-am\-ple.com}, which is an otherwise inherent and undesirable
consequence of Certificate Transparency~\cite{ct-private-subdom}. Stark et al.~\cite{stark2019ct}
reported that CAs stripping full domain names from logged certificates was a cause of breakage in
the early deployment of CT, and listed name redaction as an open problem. Eskandarian et
al.~\cite{eskandarian2017certificate} proposed a solution to supporting private subdomains based on
cryptographic commitments, but it would require updating browsers, log servers, and CA software.
Scheitle et al.~\cite{scheitle2018rise} also showed that CT logs are being actively monitored to
find new domain names as targets.

\subsubsection*{Use Case 3: Dynamic Security Policies}

One of the benefits of proxy certificates is that they lend themselves to short validity periods.
However, a more general advantage is that they allow domain owners to define security policies
dynamically and for each subdomain separately: every time a proxy certificate is created or renewed,
the domain owner is free to define a new policy or change an existing one using the appropriate
fields in the proxy certificate. We do not attempt to exhaustively list all such policies here, but
discuss a few salient examples.
The \texttt{notBefore} and \texttt{notAfter} fields define the validity period of the certificate,
while \texttt{pathLenConstraint} restricts the length of the certification path.

\looseness=-1
As we have seen, the assumption that the domain name and servers are controlled by the same entity
no longer holds. As a consequence,  a domain owner may not approve of the server's
configuration. For example, a CDN (or attacker having compromised the domain's private key) may try
to use session resumption to extend a TLS session far beyond the certificate's expiration date (see
Section~\ref{sec:tls_consequences}). With a dedicated certificate extension field in the proxy
certificate,
a security-focused domain owner could indicate to the browser that session resumption should not be
attempted. The browser may be configured to strengthen certain domain policies, but should not
weaken them.

\looseness=-1
Another example of a policy relates to how browsers deal with protocol errors. As pointed out by
Szalachowski et al.~\cite{szalachowski2014policert}, domain owners should be able to influence the
decision of the browser to either completely stop a TLS communication in case an anomaly occurs, a
\emph{hard fail}, or give the user an option to proceed despite the risk, a \emph{soft fail}.
Indeed, the domain owner is the entity best able to specify the level of security and availability
that their application requires and determine whether the browser should hard fail or soft fail. We
suggest that domain owners should be able to express their preference for a failure mode through an
extension in the proxy certificates they issue.

\section{Delegated Credentials}
\label{sec:del_cred}

A recent alternative to proxy certificates are delegated credentials, which we describe here to
enable our detailed comparison later. Similarly to proxy certificates, delegated credentials are
designed to be short-lived and allow delegation to be done offline. In contrast to proxy
certificates delegated credentials are not full-fledged X.509 certificates. A delegated credential
is composed of a public key, a validity time (relative to the certificate's \texttt{notBefore}
field), a signature algorithm, and the signature itself. In addition, every delegated credential is
bound to a signature algorithm that may be used during the TLS handshake~\cite{draft_subcerts}.

An argument put forth by the authors of the Internet Draft for using delegated credentials rather
than proxy certificates is that their semantics are minimized to mitigate the risks of
cross-protocols attacks. Moreover, software changes required to support delegated credentials are
limited to the TLS stack and do not affect PKI code.

Delegated credentials almost exclusively address the Use Case 1 of proxy certificates, that is,
delegation to content delivery networks. They do not allow domain owners to enforce security
policies on a per-subdomain basis and do not address the problem of keeping subdomains private.

\subsection{Server and Client Authentication}

Although TLS is most commonly used for server authentication, it supports client authentication as
well, and so do delegated credentials. Below we only describe server authentication but client
authentication is handled analogously.

A client supporting delegated credentials must first indicate it by sending an empty ``delegated
credential'' extension in its ClientHello message (during the TLS handshake). The server may then
send a delegated credential to the client, only if this extension is present. This mechanism ensures
that an incremental deployment is possible: delegated credentials are only provided to browsers that
support them while a less efficient alternative, such as Keyless SSL, can be used with legacy
clients. A similar mechanism could be used for an incremental deployment of proxy certificates.

\subsection{Validation and Requirements}
\label{sec:del_cred:req}

A delegated credential is validated as follows.
First, the current time must be within the validity period of the credential, and the lifetime of
the credential cannot be greater than 7 days.
Second, the signature algorithm used to sign the TLS handshake must match the one indicated in
the credential.
Finally, the end-entity certificate must satisfy the conditions we describe below, and its public
key must be used to verify the signature on the credential.

Delegated credentials introduce a new certificate extension: \texttt{DelegationUsage}. The signature
on the delegated credential may be generated using the end-entity certificate's private key, under
the conditions that the \texttt{digitalSignature} bit is set (in the key usage extension of the
certificate) and that the delegation usage extension is present. The digital signature bit is only
required when the subject public key is used for verifying signatures ``other than signatures on
certificates''~\cite{rfc5280}. Therefore, the digital signature bit is not required for validating
proxy certificates.

\subsection{Deployment Incentives}

The deployment of a new technology is always challenging, especially when it affects core Internet
protocols. On the one hand, new security policies (e.g., requiring that the server provide SCTs to
support Certificate Transparency) are often not enforced due to fear of breakage~\cite{stark2019ct}.
On the other hand, \emph{not} supporting a new scheme and thus causing breakage is different. Proxy
certificates and delegated credentials both fall into the second category, as a connection to a
server employing one of those delegation techniques might fail if the browser does not support them.

\section{TLS with Short-Lived Certificates}
\label{sec:tls_consequences}

TLS was not designed with short-lived certificates or credentials in mind. In this section, we
explain why it is critical that the certificate lifetime be considered even after a connection is
established. We stress that this issue does not only concern proxy certificates. Our discussion
concerns any TLS session that is sufficiently long-lived to extend beyond the certificate's
expiration time. This situation is not as unlikely as it may initially appear, when we consider
session resumption, but has unfortunately often been neglected in previous work on short-lived
certificates~\cite{topalovic2012towards,draft_saag_star}.

\subsection{TLS sessions and resumption}
We focus on TLS 1.3~\cite{RFC8446} and use its nomenclature, rather than that of older TLS versions.
However, similar arguments apply to older TLS versions, in particular, TLS 1.2.

TLS describes a key-exchange protocol between two parties, resulting in shared keying material used
in a subsequent \emph{connection} transferring data. Such a connection usually times out after not
being used between 1 to 5 minutes, but can otherwise stay alive indefinitely. This connection is
part of a \emph{session} that may (and in practice does) provide a \emph{pre-shared key} (PSK) which
is single-use and provisioned for session resumption, which creates a new connection.

Based on a PSK, a client can reconnect to the server while the PSK stays valid, without the
certificate being checked again. Note that a PSK stays valid for up to 7 days, at the issuing
server's discretion. Using PSK-based resumption (usually) provisions a new PSK, again valid for 7
days from its issuance, allowing indefinite chaining~\cite[Section 4.6.1]{RFC8446}.

\subsection{The problem with resumption}
At no point after the initial full key exchange is the certificate, or its lifetime, considered. The
TLS 1.3 specification suggests that the validity of connections and sessions should consider
certificate validity, but does not mandate it~\cite[at the end of Section 4.6.1]{RFC8446}. Using
session resumption can therefore nullify the security benefits of short-lived certificates. Current
web browsers do not check that a resumption is performed within the certificates' lifetime. A
malicious edge server can issue a session PSK with a long lifetime (maximum of 7 days) to clients
and prolong them on every subsequent connection, thereby subverting the benefit of short-lived
certificates.

The Internet Draft describing delegated credentials~\cite{draft_subcerts} only states that ``if a
client decides to cache the certificate chain and re-validate it when resuming a connection, the
client should also cache the associated delegated credential and re-validate it.''

\subsection{Possible solutions}
Currently, the simplest solution to this problem is to disallow session resumption completely,
thereby forsaking its efficiency benefits. For non security-sensitive websites, one can opt for
more efficiency and use longer-lived certificates together with session resumption. Unfortunately,
most brows\-ers do not allow users to configure the use of session resumption. However, some
privacy-aware browsers such as Tor Browser or JonDoBrowser disable session resumption globally 
to prevent user tracking~\cite{sy2018tracking}. We propose a more flexible approach based on dynamic
policies (as explained in Use Case 3 above, Section~\ref{sec:use_cases}) where a domain can
determine an on-off session resumption policy, which is then enforced by the browser. 

\looseness=-1
A further improvement that combines security and efficiency would require a coordination between the
lifetimes of the short-lived certificates and the session PSKs, ensuring that a PSK cannot be used
after the expiration of the edge server's (or subdomain's) certificate. For the certificate obtained
from the server during the first connection of a session, this could be achieved by dropping the
connection when the certificate expires. Unfortunately, handling the extended expiration
time of subsequently issued short-lived certificates appears hard to achieve, since it additionally
mandates a timely update of these certificates in the browser. Such an update could be achieved by
either (a) establishing a fresh session, which involves a full TLS handshake, or (b) updating the
certificate in the web browser by an out-of-band communication with the domain. The former solution
means that the use of the session PSK is limited by the lifetime of a single short-lived
certificate, which makes resumption useless with minute-range certificate lifetimes. The latter
would effectively enable the use of session resumption as long as the browser possesses a valid
certificate for the domain. However, implementing this solution would likely involve considerable
modifications to current infrastructure.

Using proxy certificates, a domain owner can specify its own session resumption policies, which are
then enforced by the browser. This allows a fine-grained split, for example by subdomain, where
security-critical subdomains (\eg a login page) disallows resumption, while other pages allow it. In
conclusion, as proxy certificates provide domain owners fine-grained control over their policies,
they achieve the efficiency benefits of quick session resumption without losing the security
benefits across all subdomains.

\section{Analysis}
\label{sec:analysis}

We now present our framework for characterizing delegation and revocation schemes with respect to a
wide range of properties. Our results are summarized in Table~\ref{tab:comparison}. We formulate the
properties in terms of the benefits that the different schemes may provide and we
categorize these benefits into six classes: revocation-related, delegation-related, security,
efficiency, deployability, and cross-category benefits. We consider 19 possible benefits in total,
two of which cover the requirements we have set out in Section~\ref{sec:introduction}. We classify
each scheme as to whether it provides, partially provides, or does not provide each
benefit.

At the bottom of the table we list various combinations of schemes that add benefits.
Short-lived proxy certificates and delegated credentials offer already most of the benefits. By
combining proxy certificates of delegated credentials with different revocation schemes, we can
achieve additional benefits.

\subsection*{Evaluation Criteria}

\newcommand\eatpunct[1]{}

\subsubsection*{A. Revocation-Related Benefits\eatpunct}~\\*[1.5ex]
\definition{Supports CA revocation}
A revocation scheme should give authorized entities (such as CAs or software vendors) the ability to
invalidate the certificates of root and intermediate CAs. We give partial points to schemes that
theoretically support revoking CA certificates but require contacting those CAs.

\definition{Supports damage-free CA revocation}
Revoking a CA certificate should not cause collateral damage, \ie it should not invalidate
certificates issued by the revoked CA before it was compromised. Standard revocation invalidates all
previously issued certificates by a revoked CA, thus leading to collateral damage.

\definition{Supports leaf revocation}
This benefit is offered by schemes that enable the revocation of certificates or credentials at the
end of the certification path. We give partial points to schemes that only propagate a manually
selected subset of revocations. Short-lived certificates and delegated credentials offer this
benefit (although, strictly speaking, they do not support revocation) because they allow domain
owners to rapidly make a compromised key unusable for the attacker, which is equivalent. Self-signed
certificates do not offer this benefit as their revocation requires additional mechanisms. Proxy
certificates get partial points as they are not necessarily short-lived.

\definition{Supports autonomous revocation}
Domain owners can autonomously (i.e., without reliance on a CA, browser vendor, or log)
perform ``leaf revocation'' (see above). This benefit corresponds to Requirement~\ref{req:revoke}.

\subsubsection*{B. Delegation-Related Benefits\eatpunct}~\\*[1.5ex]
\definition{Supports delegation}
Schemes offering this benefit allow domain owners to transfer some or all of their privileges (not
necessarily in a secure way).

\definition{Supports delegation without key sharing}
This benefit is offered by schemes that let domain owners autonomously delegate certain rights
without sharing any private key. This corresponds to Requirement~\ref{req:delegate}.

\newcommand{\ys}{{$\vcenter{\hbox{\scriptsize\CIRCLE}}$}\xspace} 
\newcommand{\no}{{$\vcenter{\hbox{\scriptsize\Circle}}$}\xspace} 
\newcommand{\pt}{{$\vcenter{\hbox{\scriptsize\LEFTcircle}}$}\xspace} 
\newcommand{\na}{--\xspace} 
\newcommand{\itm}[1]{\makebox[1em][l]{\textit{#1.}}}

\begin{table*}
	\caption{Comparison of revocation and delegation approaches. Columns represent benefits. Each
	benefit falls under one of six categories (A--F). Benefits corresponding to the requirements are
	marked with \ref{req:revoke} and \ref{req:delegate} . See
	Section~\ref{sec:background} for our categorization of revocation schemes (Cat. I--IV) and
	Section~\ref{sec:analysis} for the definitions of our criteria.}
    \centering
    \small
    \begin{tabular}{@{} c | l | c |
    	*{4}{c@{\hspace{.5em}}} |
    	*{2}{c@{\hspace{.5em}}} |
    	*{3}{c@{\hspace{.5em}}} |
    	*{3}{c@{\hspace{.5em}}} |
    	*{6}{c@{\hspace{.5em}}} |
    	*{1}{c@{\hspace{.5em}}} }
    	
        & & &

        \rotatebox[origin=lB]{90}{\textbf{Supports CA revocation}} &
        \rotatebox[origin=lB]{90}{\textbf{Supports damage-free CA rev.}} &
        \rotatebox[origin=lB]{90}{\textbf{Supports leaf revocation}} &
        \rotatebox[origin=lB]{90}{\textbf{Supports autonomous rev. (\ref{req:revoke})}} &

        \rotatebox[origin=lB]{90}{\textbf{Supports delegation}} &
        \rotatebox[origin=lB]{90}{\textbf{Delegation w/o key sharing (\ref{req:delegate})}} &

        \rotatebox[origin=lB]{90}{\textbf{Supports domain-based policies}} &
        \rotatebox[origin=lB]{90}{\textbf{No trust-on-first-use required}} &
        \rotatebox[origin=lB]{90}{\textbf{Preserves user privacy}} &

        \rotatebox[origin=lB]{90}{\textbf{Does not increase page-load delay}} &
        \rotatebox[origin=lB]{90}{\textbf{Low burden on CAs}} &
        \rotatebox[origin=lB]{90}{\textbf{Reasonable logging overhead}} &
        
        \rotatebox[origin=lB]{90}{\textbf{Non-proprietary}} &
        \rotatebox[origin=lB]{90}{\textbf{No special hardware required}} &
        \rotatebox[origin=lB]{90}{\textbf{No extra CA involvement}} &
        \rotatebox[origin=lB]{90}{\textbf{No browser-vendor involvement}} &
        \rotatebox[origin=lB]{90}{\textbf{Server compatible}} &
        \rotatebox[origin=lB]{90}{\textbf{Browser compatible}} &
        
        \rotatebox[origin=lB]{90}{\textbf{No out-of-band communication}} \\
        
        & \multicolumn{1}{c|}{Scheme} & \multirow{-4}{*}{\rotatebox[origin=lB]{90}{Reference}} &
        \multicolumn{4}{c|}{A} &
        \multicolumn{2}{c|}{B} &        
        \multicolumn{3}{c|}{C} &
        \multicolumn{3}{c|}{D} &
        \multicolumn{6}{c|}{E} &
        \multicolumn{1}{c}{F} \\
        
        \hlineB{2}
        \rule{0pt}{2.5ex}
        
        \multirow{9}{54pt}{\centering Revocation Schemes}
        & \itm{a} Regular CRL (Cat. I) & \cite{rfc5280}                    & \pt & \no & \ys & \no & \no & \no & \no & \ys & \no & \no & \no & \ys & \ys & \ys & \no & \ys & \ys & \pt & \no \\
        & \itm{b} Hard-fail OCSP (Cat. I) & \cite{rfc2560}                 & \pt & \no & \ys & \no & \no & \no & \no & \ys & \no & \no & \no & \ys & \ys & \ys & \no & \ys & \ys & \pt & \no \\
        & \itm{c} OCSP stapling (Cat. II) & \cite{rfc6961}                 & \pt & \no & \ys & \no & \no & \no & \no & \ys & \ys & \ys & \no & \ys & \ys & \ys & \no & \ys & \pt & \pt & \ys \\
        & \itm{d} PKISN (Cat. II) & \cite{PKISN2016}                       & \ys & \ys & \ys & \no & \no & \no & \no & \ys & \ys & \ys & \ys & \ys & \ys & \ys & \pt & \pt & \no & \no & \ys \\
        & \itm{e} CRLSets (Cat. III) & \cite{langley2012}                  & \ys & \no & \pt & \no & \no & \no & \no & \ys & \ys & \ys & \ys & \ys & \no & \ys & \ys & \no & \ys & \pt & \ys \\
        & \itm{f} OneCRL (Cat. III) & \cite{mozilla_rev}                   & \ys & \no & \pt & \no & \no & \no & \no & \ys & \ys & \ys & \ys & \ys & \no & \ys & \ys & \no & \ys & \pt & \ys \\
        & \itm{g} CRLite (Cat. III) & \cite{larisch2017crlite}             & \ys & \no & \ys & \no & \no & \no & \no & \ys & \ys & \ys & \ys & \ys & \ys & \ys & \ys & \no & \ys & \no & \ys \\
        & \itm{h} RevCast (Cat. IV) & \cite{schulman2014revcast}           & \ys & \no & \ys & \no & \no & \no & \no & \ys & \ys & \ys & \ys & \ys & \ys & \no & \no & \ys & \ys & \no & \ys \\
        & \itm{i} RITM (Cat. IV) & \cite{RITM2016}                         & \ys & \no & \ys & \no & \no & \no & \no & \ys & \ys & \ys & \ys & \ys & \ys & \pt & \no & \ys & \ys & \no & \ys \\[.5ex]
        
        \hline
        \rule{0pt}{2.5ex}
        
        \multirow{5}{54pt}{\centering Delegation Schemes}
        & \itm{j} SSL splitting & \cite{LesniewskiLaas:2005ib}             & \no & \no & \no & \no & \ys & \ys & \no & \ys & \ys & \no & \ys & \ys & \ys & \ys & \ys & \ys & \no & \ys & \ys \\
        & \itm{k} Keyless SSL & \cite{Bhargavan:2017dx}                    & \no & \no & \no & \no & \ys & \ys & \no & \ys & \ys & \no & \ys & \ys & \ys & \ys & \ys & \ys & \no & \ys & \ys \\
        & \itm{l} Key sharing & \cite{Cangialosi:2016jx}                   & \no & \no & \no & \no & \ys & \no & \no & \ys & \ys & \ys & \ys & \ys & \ys & \ys & \ys & \ys & \ys & \ys & \ys \\
        & \itm{m} DANE-based delegation & \cite{Liang:2014im}              & \no & \no & \no & \no & \ys & \ys & \no & \ys & \ys & \no & \ys & \ys & \ys & \ys & \ys & \ys & \ys & \no & \ys \\
        & \itm{n} Delegated credentials & \cite{draft_subcerts}            & \no & \no & \ys & \ys & \ys & \ys & \pt & \ys & \ys & \ys & \ys & \ys & \ys & \ys & \pt & \ys & \no & \no & \ys \\[.5ex]
        
        \hline
        \rule{0pt}{2.5ex}
        
        \multirow{5}{54pt}{\centering Certificate Features}
        & \itm{o} Self-signed certificates & \cite{rfc6818}                & \no & \no & \no & \no & \ys & \ys & \ys & \pt & \ys & \ys & \ys & \ys & \ys & \ys & \ys & \ys & \ys & \pt & \pt \\
        & \itm{p} Short-lived certificates & \cite{topalovic2012towards}   & \no & \no & \ys & \no & \no & \no & \no & \ys & \ys & \ys & \no & \no & \ys & \ys & \no & \ys & \ys & \ys & \ys \\
        & \itm{q} Name-constrained cert. & \cite{rfc4158}                  & \no & \no & \no & \no & \ys & \ys & \ys & \ys & \ys & \ys & \ys & \no & \ys & \ys & \no & \ys & \ys & \pt & \ys \\
        & \itm{r} Cruise-liner certificates & \cite{Cangialosi:2016jx}     & \no & \no & \no & \no & \ys & \no & \no & \ys & \ys & \ys & \ys & \ys & \ys & \ys & \no & \ys & \ys & \ys & \ys \\
        & \itm{s} Proxy certificates & \cite{welch2004x}                   & \no & \no & \pt & \pt & \ys & \ys & \ys & \ys & \ys & \ys & \ys & \ys & \ys & \ys & \ys & \ys & \pt & \no & \ys \\[.5ex]
        
        \hline
        \rule{0pt}{2.5ex}
        
        \multirow{7}{54pt}{\centering Combinations of Schemes}
        & $n + d$ &                                                        & \ys & \ys & \ys & \ys & \ys & \ys & \no & \ys & \ys & \ys & \ys & \ys & \ys & \ys & \pt & \pt & \no & \no & \ys \\
        & $n + \text{(}e\;\text{or}\:f\text{)}$ &                          & \ys & \no & \ys & \ys & \ys & \ys & \no & \ys & \ys & \ys & \ys & \ys & \no & \ys & \pt & \no & \no & \no & \ys \\
        & $n + g$ &                                                        & \ys & \no & \ys & \ys & \ys & \ys & \no & \ys & \ys & \ys & \ys & \ys & \ys & \ys & \pt & \no & \no & \no & \ys \\
        & $p + s$ &                                                        & \no & \no & \ys & \ys & \ys & \ys & \ys & \ys & \ys & \ys & \ys & \ys & \ys & \ys & \ys & \ys & \pt & \no & \ys \\
        & $p + s + d$ &                                                    & \ys & \ys & \ys & \ys & \ys & \ys & \ys & \ys & \ys & \ys & \ys & \ys & \ys & \ys & \pt & \pt & \no & \no & \ys \\
        & $p + s + \text{(}e\;\text{or}\:f\text{)}$ &                      & \ys & \no & \ys & \ys & \ys & \ys & \ys & \ys & \ys & \ys & \ys & \ys & \no & \ys & \ys & \no & \pt & \no & \ys \\
        & $p + s + g$ &                                                    & \ys & \no & \ys & \ys & \ys & \ys & \ys & \ys & \ys & \ys & \ys & \ys & \ys & \ys & \ys & \no & \pt & \no & \ys \\[.5ex]

    \end{tabular}
    \\[2ex]
    \ys = offers the benefit; \pt = partially offers the benefit; \no = does not offer the benefit.
    \label{tab:comparison}
\end{table*}

\subsubsection*{C. Security Benefits\eatpunct}~\\*[1.5ex]
\definition{Supports domain-based policies}
This benefit is offered by schemes that let domain owners define policies, \eg specifying
whether session resumption is authorized or not (see Use Case 3 in Section~\ref{sec:use_cases}).
We give partial points to delegated credentials because their semantics are limited and only
allow the domain owner to define the validity period.

\definition{No trust-on-first-use required}
This benefit is offered by schemes that do not require trusting a public key the first
time it is encountered. We give partial points to self-signed certificates as they can
be obtained through an authentic channel beforehand.

\definition{Preserves user privacy}
Browsers should not have to reveal any domain-related data to a third party when validating a
certificate or delegation.

\subsubsection*{D. Efficiency Benefits\eatpunct}~\\*[1.5ex]
\definition{Does not increase page-load delay}
This benefit is offered by schemes that do not substantially increase the time it takes for the
browser to load pages. We ignore small processing delays, but do not grant this
benefit to schemes that incur additional network delay.

\definition{Low burden on CAs}
Experience has shown that if a scheme requires CAs to make considerable operational efforts, but
provides limited financial benefits, then the scheme is unlikely to be successful. In particular,
CAs should not have to handle high numbers of requests from clients (\eg to check a revocation
status) or be required to reissue certificates at a high frequency. 

\definition{Reasonable logging overhead}
The scheme should not put excessive pressure on certificate logs. Logging certificates with a short
validity period or issued by the domain owners themselves (hence, potentially too many) would make
logs blow up. 

\subsubsection*{E. Deployability Benefits\eatpunct}~\\*[1.5ex]
\definition{Non-proprietary}
A scheme is more generally useful if it is not bound to or controlled by a particular software
vendor. This benefit is offered by schemes that are open, not controlled by a single organization,
and not restricted to a single browser.

\definition{No special hardware required}
This holds for schemes without special hardware requirements. We give partial points to schemes
that do not require  special hardware but require that existing hardware (\eg middleboxes) be
updated to support the scheme.

\definition{No extra CA involvement}
This benefit is offered by schemes that work without the participation of CAs (beyond regular
certificate issuance). We give
partial points to delegated credentials as they require the CA to issue an end-entity certificate
with special extensions (see Section~\ref{sec:del_cred:req}). 
We also give partial points to schemes that permit CA involvement but do not require it. CRLs and
OCSP do not provide this benefit as CAs typically act as issuer/responder,
although in theory this role can be fulfilled by another entity.

\definition{No browser-vendor involvement}
This benefit is offered by schemes that do not require the participation of browser vendors
beyond the need to develop compatible software. For example, browser-vendor involvement can
consist in regularly selecting and propagating a set of revocations through browser updates.
We give partial points to schemes that permit software-vendor involvement but do not require it.

\definition{Server compatible}
This benefit means that no changes are required on the server side. OCSP stapling partially offers
the benefit as it requires actions from the web server but is widely supported. Proxy certificates
also partially offer the benefit as they could be used a priori by any web server, but software
updates would be needed to let domain owners fully take advantage of proxy certificates.

\definition{Browser compatible}
This indicates that no changes to the browser are required. We give partial points to standardized
schemes that are implemented by major browser vendors but often turned off by default or improperly
enforced. Self-signed certificates also received partial points for this benefit as they typically
generate error messages on first use. We also give partial points to schemes that are specific to a
particular browser (e.g., CRLSets for Chrome only).

\smallskip
We purposely did not include ``incrementally deployable'' in our analysis as we believe that it is a
vague and potentially misleading ``benefit'' in this context. OCSP stapling, for example, can be
considered incrementally deployable in its soft-fail variant (i.e., the browser does not return an
error when the server does not support stapling), but this allows an attacker to simply omit the
OCSP status of a revoked certificate and thus provides no added security. Incremental deployment can
also be interpreted as ``clients that have not been updated to support a scheme can still
communicate with servers that support it (although the clients will not reap the security benefits
of that scheme)'', but this is captured by our ``browser compatible'' benefit. The flip side of
deploying a security scheme that is compatible with previous browser versions is that users with
outdated software are unlikely to realize that their browser does not support the latest security
standards, which does not encourage adoption.

\subsubsection*{F. Cross-Category Benefit\eatpunct}~\\*[1.5ex]
\definition{No out-of-band communication}
Requiring users to communicate with a third party or use a different channel is problematic. It can
compromise user privacy, increase latency, and captive portals may not allow the connection to be
made. We grant this benefit to schemes that require users to neither use a separate channel nor
establish a connection with a server that would not be contacted otherwise. We grant partial points
to self-signed certificates for this benefit, as an out-of-band communication may be used to
circumvent the TOFU problem self-signed certificates inherently have otherwise.

\subsection*{Combinations of Schemes}

In Table~\ref{tab:comparison}, we evaluate the benefits of combining proxy certificates and
delegated credentials with other schemes. When proxy certificates are short-lived (\ie when schemes
$p$ and $s$ are combined), revocation becomes feasible thus offering the ``supports leaf
revocation'' benefit. In that case, revocation would be undertaken by the domain owner rather than
the CA. The owner issues short-lived certificates (\eg a few hours) and simply refrains from
renewing these on time of expiry when revocation is needed.

The revocation of CA certificates is supported neither by proxy certificates nor by delegated
credentials. CA certificate revocation has a vastly different scale (with regard to the total number
of certificates) and different requirements. Therefore, we suggest that schemes that were specially
designed for that purpose (such as PKISN~\cite{PKISN2016}) or schemes that work well with smaller
numbers of certificates (such as CRLSets and OneCRL) be used in conjunction with proxy certificates
or delegated credentials.

As we could not include all possible combinations in our comparison table, we only included schemes
that offer complementary benefits with minimal drawbacks, and schemes that are already widely
deployed (CRLSets, OneCRL). We see that several combinations fulfill both of the requirements, in
most cases requiring only minor software updates.

\section{Related Work}
\label{sec:related}

\looseness=-1
In 2013, Clark and van Oorschot~\cite{DBLP:conf/sp/ClarkO13} surveyed problems of the TLS/HTTPS
ecosystem and its trust model. Their systematization of knowledge covers a wide range of problems,
including TLS protocol flaws, certification, trust anchoring, and user interface issues. They
evaluate various enhancements with respect to security properties in three categories (detecting
MitM attacks, TLS stripping, and PKI improvements) and their general impact on security and privacy,
deployability, and usability.\footnote{Some  evaluation criteria semantically overlap with theirs,
\eg ``preserves user privacy'', ``server compatible'', and ``no out-of-band communication''. Other
criteria are problem-specific.} Their analysis highlighted how CA infrastructures are increasingly
being seen as a fundamental weakness in the PKI system. The authors also analyzed techniques against
fraudulent certificates, including certificate pinning. With support for certificate pinning dropped
after years of operation~\cite{go2017pinning}, it is becoming clearer that such directions provide
little hope to address our current challenges. In contrast, our work focuses on delegation and
revocation and considers recent developments of the HTTPS ecosystem, such as delegated credentials,
Certificate Transparency, CDNs, and session resumption. Moreover, the work of Clark and van Oorschot
does not include proxy certificates and recently proposed revocation schemes. 

We now mention related topics that are relevant to the area of the
web PKI, delegation, or more generally SSL/TLS, but go beyond the scope of our study.

\subsection{Holistic PKI Proposals}

\looseness=-1
AKI~\cite{KimHuaPerJacGli13} and its successor ARPKI~\cite{BCKPSS2016} are more holistic approaches
to upgrading the web PKI. One of the main ideas in these proposals is that resilience to compromise
can be improved by requiring that multiple CAs sign each end-entity certificate. Additionally, to
guarantee that no other rogue certificate exists for a given domain, all certificates must be logged
and log servers  efficiently produce both presence and absence proofs. Unfortunately, this also
implies that only one certificate per domain can be valid at any given time. ARPKI's key security
properties were also formally verified.

PoliCert~\cite{szalachowski2014policert} builds on top of ARPKI to solve the unique-certificate
problem by replacing the end-entity certificate by a unique domain policy that specifies which CAs
are allowed to issue (potentially multiple) certificates for that domain. However, that approach
does not allow domain owners to rapidly change their policies or produce their own certificates (\ie
without contacting several CAs). Therefore, proxy certificates could complement ARPKI certificate
chains as a lightweight and more dynamic alternative to PoliCert.

\subsection{Beyond HTTPS Delegation}

A number of previous research papers have addressed the problem of delegation in different contexts
than that of the web. Kogan et al.~\cite{Kogan:2017bf} argue that a secure delegation system should
always make explicit ``\emph{who} will do \emph{what} to \emph{whom}'', and present a design for the
SSH protocol, called Guardian Agent. MacKenzie et al.~\cite{MacKenzie:2003gg} address the problem of
server delegation in the context of capture-resilient devices (\ie devices required to confirm
password guesses with a designated remote server before private-key operations).
STYX~\cite{wei2017styx} is a key management scheme, based on Intel SGX, Intel QuickAssist
Technology, and the SIGMA (SIGn-and-MAc) protocol, which can be used to distribute and protect
SSL/TLS keys.

\subsection{Other Issues with SSL/TLS}

Sy et al.~\cite{sy2018tracking} recently showed, after analyzing as many as 48 browsers, that
session resumption was also problematic for user privacy as it can be used to track the average user
for up to eight days with standard settings. With a long session resumption lifetime, a majority of
users can even be tracked permanently. Problems have also been discovered in the way CAs perform
domain validation: exploiting a BGP vulnerability to hijack traffic, an attacker can obtain a rogue
certificate from vulnerable CAs~\cite{birge18bamboozling,brandt18domain}. Brubaker et al. also found
serious vulnerabilities in popular implementations of SSL/TLS using certificates with unusual
combinations of extensions and constraints~\cite{Brubaker:2014:UFA:2650286.2650793}.

\section{Conclusion}
\label{sec:conclusion}

Delegation and revocation sometimes appear to be two sides of the same coin. We have shown that it
is possible to simultaneously address both of these issues with a slight modification to the chain
of trust, but this is a delicate task. On the one hand, using a single long-lived certificate per
domain has several advantages: it is more practical for administrators, it lowers the pressure on
transparency logs, it reduces CA-incurred costs, and thus benefits the deployment of HTTPS. On the
other hand, using short-lived credentials, not sharing private keys with third parties, and limiting
the scope and exposure of private keys is preferable from a security standpoint.

Delegated credentials and proxy certificates solve this conundrum with minor changes to existing
standards and little overhead. They give more flexibility to domain owners and accommodate practices
that have become ubiquitous on the web, such as delegation to CDNs. Although proxy certificates and
delegated credentials address similar issues, proxy certificates offer more flexibility. For
example, they enable domain owners to maintain as many certificates as they desire, without CA
reliance, with policies and validity periods defined individually for each subdomain. Proxy
certificates also address the problem of keeping private subdomains hidden from certificate logs. 

We thus view the recent deployment efforts surrounding delegated credentials with optimism, but note
that many challenges remain. For example, session resumption---whose goal is to bring page-load
delay to a minimum, which in turn is one of the main reasons for using a CDN in the first
place---finds itself in conflict with short-lived credentials. Also, the revocation of CA
certificates is a major issue that short-lived credentials do not address. Other schemes bring
solutions to these issues but combining them in meaningful ways is hard. Some of the examined
proposals may be conceptually simple, but have considerable ramifications in the highly complex
HTTPS ecosystem. We hope that our systematic analysis sheds light on these issues and helps guide
future research.

\bibliographystyle{plain}
\bibliography{references.bib}

\end{document}